\documentclass[proof]{pasj00}
\draft

\begin{document}
\SetRunningHead{H. Nakanishi and Y. Sofue}{3D Distribution of H {\sc i} gas in
the Milky Way Galaxy}
\Received{2002/05/31}
\Accepted{2002/11/08}

\title{Three-dimensional distribution of the ISM in the Milky Way
Galaxy: I. The H {\sc i} Disk}

\author{Hiroyuki \textsc{Nakanishi} and Yoshiaki \textsc{Sofue}}%
\affil{Institute of Astronomy, the University of Tokyo, 2-21-1 Osawa, Mitaka,
Tokyo 181-0015}
\email{hnakanis@ioa.s.u-tokyo.ac.jp}


%


\maketitle

\begin{abstract}
We derived the three-dimensional distribution of H {\sc i} gas in the
 Milky Way Galaxy using the latest H {\sc i} survey data cubes and
 rotation curves. The distance of the H {\sc i} gas was determined by 
 the kinematic distance using a rotation curve. We solved the near--far
 problem in the inner Galaxy by a fitting method which involves 
 introducing a model of vertical H {\sc i} distribution. 
 In our resultant maps we could trace three prominent arms: 
 the Sagittarius--Carina arm, the Perseus arm, and the Outer
 arm. These three arms were found to be logarithmic spiral arms. The
 pitch angles of the Sagittarius--Carina, Perseus, and 
 Outer arms were estimated to be about $11\arcdeg$, $18\arcdeg$, and $7\arcdeg$,
 respectively.  The Sun is located in a region rich in H {\sc i} gas
 between the Sagittarius--Carina arm and the Perseus arm. 
 The H {\sc i} disk shows large and asymmetric warping in the outer disk:
 the H {\sc i} disk goes up to about 1.5 kpc above the Galactic
 plane in the northern hemisphere, and down to about 1 kpc in
 the southern hemisphere, which means asymmetric warping. The 
 inner H {\sc i} disk is also found to be tilting. 
 The radius of the H {\sc i} disk is about 17 kpc and the H{\sc
 i} mass within this radius is estimated to be $2.5 \times 10^9$ M$_\odot$, which corresponds to 1.5\% of the dynamical mass predicted from the rotation curve. We also found that the H {\sc i} outskirt is largely swelling in the fourth quadrant, and hence the Galaxy is significantly lopsided. The scale-height of the H {\sc i} layer increases with the radius, and is correlated with the H {\sc i} volume density at the centroid of the H {\sc i} layer.  
\end{abstract}

\section{Introduction}
While the Milky Way Galaxy is the most explored galaxy, its global
three--dimensional (3D) structure, such as the spiral arms, has not yet been
clarified, because we are embedded in it. 
One of the best tracers of the global structure of the Galaxy is 
neutral hydrogen (H {\sc i}) gas, because H {\sc i} gas is one of the
major components of the interstellar matter (ISM) in spiral galaxies, and the
ISM is almost optically thin against the 21-cm line of the H {\sc i} gas.

The earliest study of the global structure of the Milky Way using 
H {\sc i} data was done by \citet{oor58}, who presented a face-on H {\sc i}
density map of the whole Galaxy. They reported several H{\sc
i} arms in the Milky Way: the so-called  
Sagittarius arm, the Orion arm, and the Perseus arm.
Later, as more H {\sc i} surveys were achieved, \citet{ker69},
\citet{wea70}, and \citet{ver73} traced larger-radius H {\sc i} arm,
which is the so-called Outer arm, on a 
longitude--velocity diagram. \citet{kul82} analyzed the H {\sc i} data
to make an H {\sc i} density map for the outer Galaxy, and traced two
arms having a pitch angle of $\sim 22\arcdeg$--$25\arcdeg$. 
Burton and te Lintel Hekkert (1986) showed three--dimensional H {\sc i} maps of the
outer Galaxy for all Galactic longitudes, and reported a warped H {\sc
i} disk. \citet{dip91} also presented an H {\sc i} density distribution
of the outer Galaxy, and discussed the warped disk and the flaring H {\sc i}
disk. 

More recently,  Hartmann and Burton (1997) archived a deeper and more complete
survey with a much higher sensitivity. In the present work, we tried to convert this
latest data cube into an H {\sc i} distribution. Combining with the southern
survey data cubes available, we aimed to derive a 3D map of the whole Galaxy.
This paper is the first of a series presenting our analysis of the 3D
structure of the ISM. In the second paper, we will present the result
for molecular gas. 

In this paper we describe the data and our method for deriving the H {\sc i}
distribution of the whole Galaxy in sections \ref{data} and 
\ref{method}, and report on the resultant H {\sc i} maps and some properties of
H {\sc i} disk of the Galaxy in section \ref{result}.

\section{Data}
\label{data}
\subsection{H {\sc i} Survey Data}
Three available H {\sc i} survey data are used in this paper: the
Leiden/Dwingeloo survey \citep{har97}, Parkes survey \citep{ker86}, and
NRAO survey \citep{bur83}, which all are archived in the FITS format.
We took data within a latitude range of $-10\arcdeg < b <
+10\arcdeg$ from these data cubes to study the H {\sc i} distribution.

The Leiden/Dwingeloo survey \citep{har97} is used for the
northern hemisphere, which was obtained with the 25 m
Dwingeloo radio telescope in the Netherlands with a half-power beam
width (HPBW) of $0\fdg6$.
The data consists of 21 cm line spectra over the entire sky above
declinations of $-30\arcdeg$ with a grid spacing of $0\fdg5$. The radial
velocity ($V_{\rm r}$) coverage is from $-450$ to $+400$ km s$^{-1}$
with a velocity resolution of $\simeq 1.03$ km s$^{-1}$.
The r.m.s. noise level is 0.07 K. These spectra are averaged so that the
bin width of latitude becomes $1\arcdeg$ in this paper.

The Parkes survey \citep{ker86} is used for the southern hemisphere,
which was obtained  with the CISRO 18 m telescope in
Australia with an HPBW of $0\fdg8$.
It covers Galactic longitudes from $l = 240\arcdeg$ to $350\arcdeg$ with
a grid spacing of $\Delta l=0\fdg5$, and Galactic latitude
$b=-10\arcdeg$ to $+10\arcdeg$, with a grid spacing $\Delta
b=0\fdg25$. The velocity range is 300 km s$^{-1}$ with a velocity
resolution of $\simeq 2.1$ km s$^{-1}$.
The average r.m.s. is 0.27 K. These spectra are also averaged so that the
bin width of latitude becomes $1\arcdeg$.

The Galactic center H {\sc i} survey of National Radio Astronomy
Observatory is used for the southern hemisphere between $l=350\arcdeg$
and $l=360\arcdeg$ \citep{bur83}.
This data cube was obtained with the 140-feet telescope of the NRAO, whose
HPBW is $21\arcmin$. The survey
covers the region $348\arcdeg < l < 10\arcdeg$, $-10\arcdeg < b < 10\arcdeg$, and a grid spacing of
$0\fdg5$ in both $l$ and $b$. The velocity range is $|v| < 310 $ km
s$^{-1}$ 
and the  velocity resolution 5.5 km s$^{-1}$.  The r.m.s. level is 0.03
K. These spectra are also averaged so that the bin width of latitude
becomes $1\arcdeg$.

\subsection{Rotation Curve}
Since the 21-cm spectra have only information about the radial velocity of
the H {\sc i} gas, we need to transform the radial velocity, $V_{\rm
r}$, 
of the ISM in the Galaxy into the distance, $r$, in the line of sight.
The relation between the radial velocity, $V_{\rm r}$, and the distance,
$r$, in the line of sight is obtained using the rotation curve of the
Galaxy, assuming that it rotates circularly. 

For the inner rotation curve, we use Clemens's rotation curve
\citep{cle85}, which was derived from the Massachusetts--Stoney Brook
Galactic equator CO survey. Clemens's rotation curve was fitted by a
polynomial of the form

\begin{equation}
V(R) = \Sigma_{{\it i} =0}^7 A_{\it i} R^{\it i}. 
\end{equation}

Though Clemens listed the fitted coefficients, $A_{\it i}$, for
the cases of $(R_0, V_0) = (8.5 \mbox{ kpc},220 \mbox{ km s}^{-1})$ and
$(10 \mbox{ kpc},250 \mbox{ km s}^{-1})$, we modified them
for the case of $(R_0, V_0) = (8 \mbox{ kpc},217 \mbox{ km s}^{-1})$
in order to connect smoothly to the outer rotation curve in this
paper. Modified coefficients are listed in table \ref{coefficient}.

\begin{table}
  \caption{Modified coefficients of the inner rotation curve.}\label{coefficient}
  \begin{center}
    \begin{tabular}{cccc}
     \hline\hline
      & $R < 0.09$ $R_0$ & $0.09$ $R_0 < R < 0.45$ $R_0$ & $0.45$ $R_0 < R < R_0$\\ 
     \hline 
     $A_0$ & 0. & 325.0912 & $-2342.65649$ \\
     $A_1$ & 3261.30 & $-264.0309$ & 2663.95435 \\
     $A_2$ & $-17847.8$ & 261.7602 &$-1156.07764$ \\
     $A_3$ & 52752.5 & $-132.822906$ & 269.354645 \\
     $A_4$ & $-87027.2$ &31.9537735 & $-36.2039909$ \\
     $A_5$ & 74344. & $-2.857958$ & 2.8025787 \\
     $A_6$ & $-25510.$ & 0. & $-0.1158279$ \\
     $A_7$ & 0. & 0. & 0.001977 \\
     \hline
    \end{tabular}
  \end{center}
\end{table}

For the outer rotation curve, we use a model rotation curve of 
\citet{deh98}, which was determined by fitting to
observational constraints ever obtained. However, because
of the small number of data points for the outer rotation curve, the mass
model has not been settled yet, and four types of models are presented
by \citet{deh98}. Of the four models, Model 2 is adopted in this work, 
since it has the smallest deviation from the observation among the
four. In using this rotation curve, we adopted $R_0=8$ kpc, $V_0 = 217$
km s$^{-1}$, which are used for the Galactic constant in \citet{deh98}.
The shape of the whole rotation curve used in this paper is shown in
figure 1.

\section{METHOD}
\label{method}
In this section we describe a method for transforming the 21-cm
spectra into an H {\sc i} density distribution. Due to a geometrical
reason, the H {\sc i} distribution should be calculated separately in
several cases: the outer Galaxy, the inner Galaxy, and near the tangential
points.
The radial velocity and the distance in a line of
sight can be uniquely determined in the outer Galaxy, but the distance
cannot be determined uniquely from the radial velocity in the inner Galaxy
because of the near-far ambiguity. Since the H {\sc i} gas has
a velocity dispersion, the H {\sc i} density at the tangential point
cannot be calculated without considering its effects.

\subsection{Basic Idea and Derivation of the H {\sc i} Distribution
   in the Outer Galaxy}
\label{outer}
The distance of the H {\sc i} gas having a radial velocity $V_{\rm r}$
can be calculated as follows. Assuming that the motion of the Galactic
ISM is purely circular and that the gas away from the Galactic plane rotates
at the same speed as the gas in the underlying disk, the radial velocity
$V_{\rm r}$ of the ISM at Galactic longitude $l$, Galactic latitude $b$, 
and Galactocentric distance $R$ is expressed by

\begin{equation}\label{Vr}
V_{\rm r}(l,b) = \left[ {R_0 \over R}V(R) - V_0 \right]
\sin{l} \cos{b}.
\end{equation}
The heliocentric distance $r$ of the ISM at $(R,l,b)$ is obtained by
solving the equation
\begin{equation}\label{costheory}
R^2 = r^2 + R_0^2 - 2rR_0 \cos{l}. 
\end{equation}
In the outer part of the solar orbit ($R>R_0$), we have one
solution, $r=R_0 \cos{l} + \sqrt{R^2 - R_0^2 {\rm sin}^2 l}$, 
because the other solution has a negative value, which is inadequate.

In the inner Galaxy ($R < R_0$) we have two adequate solutions, $r=R_0
\cos{l} \pm \sqrt{R^2 - R_0^2 {\rm sin}^2 l}$; that is, we have the two-fold of
kinematic distance in the inner Galaxy. However, at a tangential point
where $R=R_0 \sin{l}$, equation (\ref{costheory}) has only one solution despite the inner
Galaxy. Hence, we have no ambiguity of distance at this point, and the
kinematic distance 
is uniquely determined for a given radial velocity. For example, we plot
the radial velocity against the heliocentric distance ($r$) in the
case of $l=30\arcdeg$ in figure 2. The radial
velocity in the longitude range $l < 90\arcdeg$ becomes positive in the
inner Galaxy, and increases toward the
tangential point where it gets maximum. Beyond the tangential point it
decreases monotonically and becomes negative in the outer Galaxy. The radial
velocity for the whole Galactic equator is plotted in figure 3.

The H {\sc i} density can be estimated as follows. If the 21-cm
line is assumed to be optically thin,  the H {\sc i} column density
$N_{\rm HI}$ [cm$^{-2}$] is given by integrating brightness
temperature, $T_{\rm b}$ [K], over radial velocity, $V_{\rm r}$ [km
s$^{-1}$],
\begin{equation}
N_{\rm HI} \mbox{[cm$^{-2}$]} = 1.82 \times 10^{18}
 \int^{V_{{\rm r}1}+ \Delta V_{\rm r}}_{V_{{\rm r}1}}
T_{\rm b} dV_{\rm r}.
\end{equation}
In this work, the velocity width, $\Delta V_{\rm r}$,  
is taken to be $2.06$ km s$^{-1}$ in the case of the
Leiden/Dwingeloo survey, $2.00$ km s$^{-1}$ in the case of the
Parkes survey, and 5.5 km s$^{-1}$ in the case of the NRAO survey 
based on the velocity resolution of the southern survey.

The average volume density at an individual point is given by dividing
the column density by the path length, $\Delta r$, which contributes
emission in the velocity range $V_{{\rm r}1} < V_{\rm r} < V_{{\rm r}1}
+ \Delta V_{\rm r}$: 
\begin{equation}\label{TbtonHI}
n_{\rm HI} [{\rm cm}^{-3}] = {N_{\rm HI} \over \Delta r} = 1.82 \times 10^{18}
T_{\rm b} {\Delta V_{\rm r} \over \Delta r}.
\end{equation}

Thus, we obtain the H {\sc i} densities, $n_{\rm HI}$, in the outer Galaxy from
the 21-cm spectra $T_{\rm b}(l,b,V_{\rm r})$. However, the local gas
with a high velocity dispersion contributes to the high-altitude
emissions. Therefore, the obtained H {\sc i} distribution includes the
local gas in the form of $n_{\rm observed} = n_{\rm HI} + n_{\rm local}$. 
In this paper, we assumed that the local gas can be expressed by a linear
function of the Galactic latitude, $b$, and subtracted it from
$n_{\rm observed}$. Because of this procedure, the diffuse H {\sc i} gas,
which extends to
a high Galactic altitude \citep{dip91}, is subtracted, and can not be 
detected in the maps presented in this paper.

\subsection{H {\sc i} Density around the Tangential Points}
At a tangential point $r=R_0\cos{l}$ the kinematic distance can be
uniquely determined even for the inner Galaxy.
Since the H {\sc i} gas has an intrinsic velocity dispersion, the H {\sc
i} spectra have a higher velocity component than the terminal velocity, $V_{\rm
t}$. Velocity dispersion, $\sigma$, is assumed to be 10 km s$^{-1}$;  
\citet{mal95} estimate it to be $9.2$ km s$^{-1}$ for the first quadrant
($0\arcdeg <  l < 90\arcdeg $), $9.0$ km s$^{-1}$ for the fourth quadrant
($270\arcdeg <  l < 360\arcdeg$).
In a given line of sight, the H {\sc i} gas having a radial velocity
of the range $|V_{\rm t}| - \sigma \leq |V_{\rm r}| \leq |V_{\rm t}|$
contributes to a higher velocity component than the terminal velocity due
the velocity dispersion. Also, equation (\ref{TbtonHI}) at the
tangential points does not work because $\Delta V_{\rm r} / \Delta r$
goes to zero there. 
Therefore, the volume density of the H {\sc i} gas  giving the
radial velocity $|V_{\rm t}| - \sigma \leq |V_{\rm r}| \leq |V_{\rm t}|$
around the tangential points is calculated by integrating the emission
within the velocity range of $|V_{\rm t}| - \sigma \leq |V_{\rm r}| \leq
\infty$, as follows:  
\begin{equation}
n_{\rm HI}(r_{\rm t})=
{1.82 \times 10^{18} \int^{\infty}_{V_{\rm t}-\sigma}T_{\rm b} dV_{\rm r}
\over {r_2(V_{\rm t}-\sigma)- r_1(V_{\rm t}-\sigma)}},   \mbox{ ($0\arcdeg < l < 90\arcdeg$)}, 
\end{equation}
\begin{equation}
n_{\rm HI}(r_{\rm t})=
{1.82 \times 10^{18} \int_{-\infty}^{V_{\rm t}+\sigma}T_{\rm b} dV_{\rm r}
\over {r_2(V_{\rm t}+\sigma)- r_1(V_{\rm t}+\sigma)}},   \mbox{ ($270\arcdeg < l < 360\arcdeg$)}, 
\end{equation}
where $r_1(V_{\rm t} \pm \sigma)$ and $r_2(V_{\rm t} \pm \sigma)$ denote 
the heliocentric distances of the points giving radial velocities of $ V_{\rm r}
\pm \sigma $. Subscripts 1 and 2 indicate the near and far points,
respectively.

\subsection{Derivation of the H {\sc i} Distribution in the Inner Galaxy}
\label{inner}
In the inner Galaxy there are two points giving the same radial velocity
in a given line of sight, and thus the observed spectra give the total
emission at the near and far points. We solved this problem of the
distance-degeneracy by considering the H {\sc i} layer thickness.

First of all, we determined the H {\sc i} layer thickness and its
variations using the Galactocentric distance.
For the model of the vertical distribution of H {\sc i} we use  
Spitzer's model \citep{spi42}.
\citet{spi42} analytically solved the dynamical $z$-distribution of stars
and ISM.
Assuming that the stars and gas in the Galactic disk are isothermal and
self-gravitating, the number density $n_{\rm HI}$ can be expressed
as a function of $z$ by solving Poisson's equation in
hydro-dynamical equilibrium. We then obtain 
\begin{equation}\label{spi_eq}
n_{\rm HI} (\xi)= n_{{\rm HI}_0}  {\rm sech}^2(\xi)
\end{equation}
with $\xi$ being
\begin{equation}
\xi = \log{(1 + \sqrt{2})}{z - z_0 \over z_{1/2}},
\end{equation}
where $z$ is the height from the Galactic equator, $z_0$ is the height where
$n_{{\rm HI}}$ becomes the maximum value, $n_{{\rm HI}_0}$, and
$z_{1/2}$ is the height where $n_{\rm HI}$ becomes half the value of
the maximum, which is half the full width of the half maximum (FWHM).

In equation (\ref{spi_eq}) we assume that $n_{{\rm HI}_0}$ and $z_0$ vary with its
position, but the scale-height, $z_{1/2}$, depends on the Galactocentric
distance alone.
We determined the H {\sc i} layer thickness $z_{1/2}$ as a function of the
Galactocentric distance by fitting the model function to the observed
data.
The thickness of the inner H {\sc i} layer was determined by fitting the
model function to the vertical distribution at the tangential point
obtained in the previous section; we substituted the values into three
parameters, ($n_{{\rm HI}_0}$, $z_0$, and $z_{1/2}$) in equation
(\ref{spi_eq}), and chose the best set of parameters to fit to the
observation by the least-squares method. The steps to search for the 
parameters were $0.001$ of the peak for $n_{{\rm HI}_0}$, 1 pc for $z_0$
and $z_{1/2}$.
On the other hand, the thickness of the outer disk was determined by
fitting the model to all points, since there is no distance-ambiguity
in the outer Galaxy.  The resultant thickness of the H {\sc i} layer is
presented in figure 4.

For the next stage, the total emission was divided into two H {\sc i}
densities at the near and far points in the line of sight using the thickness
of the H {\sc i} layer.
First, consider an H {\sc i} gas having a radial velocity $V
<V_{\rm r} <V + \Delta V$ for a given longitude, $l$. This gas locates at
corresponding kinematic distances, $r_1 - \Delta r_1 /2 < r< r_1 +
\Delta r_1 /2 $, and $r_2 - \Delta r_2/2 < r< r_2 + \Delta r_2/2$.
Assuming that the H {\sc i} gas follows Spitzer's model, the H {\sc i}
column density, $N_{\rm HI}(b)$, is represented as a function of latitude
with four parameters: $n_{\rm HI_{0_1}}$, $n_{\rm HI_{0_2}}$, $z_{0_1}$,
and $z_{0_2}$. That is, 
\begin{equation}
N_{\rm HI}(b) = n_{\rm HI_{0_1}} {\rm sech}^2(\xi_1)\times {\Delta r_1
\over \cos{b}} + n_{\rm HI_{0_2}} {\rm sech}^2(\xi_2) \times
{\Delta r_2 \over \cos{b}}, 
\end{equation}
where
\begin{equation}
\xi_1 = \log{(1+\sqrt{2})}{r_1 \tan{b} - z_{0_1}\over z_{1/2}}, 
\end{equation}

\begin{equation}
\xi_2 = \log{(1+\sqrt{2})}{r_2 \tan{b} - z_{0_2}\over z_{1/2}}. 
\end{equation}

The four parameters were determined by fitting the observed
data by a model using the least-squares method. Finally, the H {\sc i}
density at the near and far points were calculated using $(n_{\rm HI_{0_1}},
z_{0_1})$ and $(n_{\rm HI_{0_2}},z_{0_2})$, respectively.

\subsection{Construction of the 3D H {\sc i} Density Map}
We thus obtain an H {\sc i} density cube, $n_{\rm HI}(l, b, r)$, from the
21-cm spectra $T_{\rm b}(l, b, V_{\rm r})$. Finally, the $n_{\rm HI}$ data
cube in the $(l, b, r)$ space was transformed into $n_{\rm HI}$ in 
$(x, y, z)$ space where $x, y, z$ are the Cartesian coordinates in the
Galaxy. The position $(x, y, z)=(0, 0, 0)$ indicates the Galactic center (G.C.), and the Sun is situated at $(0
\mbox{ kpc}, \mbox{ }8.0 \mbox{ kpc}, \mbox{ }0 \mbox{ kpc})$. The space of $(l, b, r)$
was also transformed into cylindrical coordinates $(R, \theta, z)$, where
$R$ and $\theta$ are the distance from the rotation axis and the azimuth
angle around it, respectively. The angle $\theta$ was taken so that
$\theta = 180\arcdeg$ points to the Sun and $\theta = 90\arcdeg$ is
parallel to $l = 90\arcdeg$.

\section{Results}
\label{result}
The derived H {\sc i} density distribution is presented in the
forms of (1) a face-on H {\sc i} column density map, (2) face-on H{\sc
i} volume density maps sliced at several altitudes, and (3) edge-on
H {\sc i} volume density maps sliced at several azimuthal angles around
the Galactic center.

\noindent{1. \it face-on column density map}---
Figure 5 shows a face-on column density map of H {\sc i} gas 
in the Galaxy. According to the \citet{spi42}'s model, the H {\sc i} column density, $\Sigma$, is calculated as
\begin{equation}
\label{eq-columndensity}
\Sigma (R, \theta) = 2.27 n_{{\rm HI}_0} z_{1/2}. 
\end{equation}

\noindent{2. \it face-on volume density maps}---
Figure 6 shows face-on volume density maps. Each map shows the H {\sc i} volume densities
 in a sheet sliced parallel to the Galactic equator. The interval of each
 map is 500 pc; $z=+1000$ pc, $+500$ pc, $0$ pc, $-500$ pc, and $-1000$ pc.

\noindent{3. \it edge-on volume density maps}---
Figure 7 shows edge-on volume density maps. Each map shows the
H {\sc i} volume densities in
 a sheet through the Galactic center perpendicular to the Galactic
 plane. Individual panels show sections through $\theta=50\arcdeg$ -- $230\arcdeg$, $\theta=80\arcdeg$ -- $260\arcdeg$, $\theta=110\arcdeg$ -- $290\arcdeg$, and  
 $\theta=140\arcdeg$ -- $320\arcdeg$.

\subsection{H {\sc i} Spiral Arms}
Some H {\sc i} spiral arms can be traced in the resultant face-on maps,
figure 5 and figure 6. In order to
investigate the properties of the spiral arms, the H {\sc i} column
density is shown in polar coordinates in figure 8. The horizontal and vertical
axes represent the azimuthal angle $\theta$ and the distance $R$ from the Galactic
center in logarithmic scale, respectively.  Each arm on the polar
diagram is found to be aligned in a straight line, indicating that
the arms are logarithmic spiral arms. Schematic tracers of H {\sc i} arms are
presented in figure 9.

The largest arm is the so-called Outer arm which is traced from
$(R, \mbox{ }\theta) \sim (7.0 \mbox{ kpc}, \mbox{ }-125\arcdeg)$ to $(R, \theta) \sim
(12.1 \mbox{ kpc}, \mbox{ }130\arcdeg)$. The pitch angle is estimated to be 
$\sim 7\arcdeg$. Another prominent arm is traced from
$(R, \mbox{ }\theta) \sim (5.6 \mbox{ kpc}, \mbox{ }100\arcdeg)$ to $(R,
\mbox{ }\theta) \sim (11.4  
\mbox{ kpc}, \mbox{ }310\arcdeg)$, which is the so-called Sagittarius--Carina
arm. Its pitch angle is also estimated to be $\sim 11\arcdeg$. 
Another arm can be traced from $(R, \mbox{ }\theta) \sim (5.8 \mbox{
kpc}, \mbox{ }50\arcdeg)$ to $(R, \mbox{ }\theta) \sim (13.6 \mbox{ kpc},
\mbox{ }200\arcdeg)$, which corresponds to the so-called Perseus arm. 
The pitch angle is $\sim 18\arcdeg$. This arm may apparently connect to
the Outer arm (figure 5), but has a clearly different pitch
angle. Hence, they may be distinct arms. 

The Sun is located at a relatively H {\sc i}-rich region between the
Sagittarius--Carina arm and Perseus arm, which is called the local arm. 

The Galactic spiral structure has been studied by many researchers 
using various methods. 
\citet{oor58} found three H {\sc i}
spiral arms: the Sagittarius, Orion, and Perseus arms. 
Although the Orion arm may correspond to the local arm, the others are
in good agreement with our traced arms. \citet{wea70} found the 
Outer arm, which also agrees with our result. They estimated the 
pitch angle of the Sagittarius--Carina arm to be
$12.5\arcdeg$, which is consistent with our result. \citet{kul82}
reported three spiral arms, and estimated the pitch angles to be
$22\arcdeg$ -- $25\arcdeg$, which are pretty large compared to our values. 
\citet{geo76} presented four spiral arms using the H{\sc ii} regions. They
traced the Sagittarius--Carina arm with a pitch angle $10\arcdeg$ --
$15\arcdeg$, and the Perseus arm with s pitch angle of $11\arcdeg$, which are consistent
with our results. However, their Scutum--Crux and
Norma arms could not be confirmed in our H {\sc i} map. 
\citet{dri01} fitted the far and near infrared data form COBE/DIRBE by a 
2- and 4-armed logarithmic spiral model. Our Sagittarius and Perseus
arms are consistent with their spiral structure. Also, our result that
the Galactic arms are logarithmic spiral arms confirms their
logarithmic spiral model.

\subsection{Size and Mass of the H {\sc i} Disk}
The radial distribution of the surface density of the H {\sc i} gas is shown
in figure 10 by averaging it azimuthally. The surface
density is low near the Galactic center, and is lower than 1.9 $M_\odot
{\rm pc}^{-2}$ within a radius of 6 kpc. The H {\sc i} surface
density attains the maximum values at radii 7 -- 12 kpc, where it
becomes larger than 4 $M_\odot \mbox{pc}^{-2}$.  Then, the H {\sc i} density
decreases toward the outer region. It amounts to $\sim 1$ $M_\odot
\mbox{pc}^{-2}$ at radius 17 kpc, and therefore the 
size of the H {\sc i} disk is 17 kpc according to the definition of
\citet{bro94}. This radius of the H {\sc i} disk is about 1.3--times as
large as the stellar disk, since the stellar disk is estimated to be 13
kpc assuming $R_0 = 8$ kpc \citep{rob92}. This ratio is typical for
general spiral galaxies. 

The H {\sc i} mass within a radius of 17 kpc was calculated to be $2.5 \times
10^9$ $M_\odot$, which corresponds to 1.5\% of the total mass predicted
by the rotation curve. 

\subsection{Warping and Tilting H {\sc i} Disk}
Figure 7 shows that the H {\sc i} disk is strongly warping
at the outer
region, which is consistent with former studies \citep{bur86,dip91}. The
amplitude of warping is 
large in the directions  
 $\theta = 80\arcdeg$ and $\theta = 260\arcdeg$. In the direction 
$\theta = 80\arcdeg$, warping starts near $R = 12$ kpc, and the H {\sc i}
disk is bending up to $z = + 1.5$ kpc height from the Galactic plane at $R=16$ kpc. 
Also, in the direction of $\theta = 260\arcdeg$, warping also starts at 
$R=12$ kpc. Then, the H {\sc i} disk is bending down to $z = - 1$ kpc
height from the Galactic plane at $R=16$ kpc, and is flopping back to $z
= - 0.5$ kpc height at $R \sim 18$ kpc. Thus, the Galactic warping is
found to be asymmetric. Since the size of the stellar disk is thought to
be about 13 kpc,  the warping starts at the edge of the stellar disk,
which follows one of the rules of warping \citep{bin92}. Hence, the
warping is asymmetric. 

This work displays the tilting of the inner disk \citep{bur78,san84}. 
In figure 11, the centroid of the H {\sc i} disk $z_0$ at $R=5$
kpc is plotted versus azimuthal angle $\theta$ for an instance, which shows
that the H {\sc i} disk is tilted against the $b=0\arcdeg$ plane. 
The amplitude of the tilting is $\sim 100$ pc. The H {\sc i} disk is bending up in the direction $\theta \sim 80\arcdeg$,
and is bending down in the direction $\theta \sim 260\arcdeg$, which is the
same trend as the warping of the outer disk. 


\subsection{Lopsided H {\sc i} Disk}
Figure 5 shows that the Galactic H {\sc i} disk is not symmetric. The
H {\sc i} disk is swelled in the direction $\theta = 310\arcdeg$ in
the third quadrant. The radius of the outer H {\sc i} disk (defined by the
contours at $8.0 \times 10^{19}$ H cm$^{-2}$ in the second quadrant) is
$\sim 11$ kpc, whereas it is $\sim 22$ kpc in the fourth 
quadrant. Hence, the Galaxy is significantly lopsided in H {\sc i}.

\subsection{Scale-Height of the H {\sc i} Disk}
We define the scale-height of the H {\sc i} disk as the value of $z_{1/2}$
(see subsection \ref{inner}). The azimuthally averaged scale-height is
plotted in figure 4 versus the radius. Figure 4 shows that the
scale-height increases 
with radius, from $\sim 100$ pc at $R=1$ kpc to $\sim 700$ pc at $R=17$ kpc.  
According to the \citet{spi42}'s model, the scale-height is calculated as 
\begin{equation}
\label{eq-scaleheight}
z_{1/2} = 0.274 \sqrt[]{{2 \sigma^2 \over 3 \pi G \rho}}. 
\end{equation}
Assuming that velocity dispersion is constant at any radius, the
increase in the scale-height implies a decrease in the
surface mass density at the Galactic plane. 

In subsection \ref{outer}, the volume density of H {\sc i} gas ($n_{{\rm HI}0}$) and
scale-height ($z_{1/2}$) are obtained for each point of the outer Galaxy. 
In figure 12, we plot the scale-height, $z_{1/2}$, versus H {\sc i} volume
density, $n_{{\rm HI}0}$, in an annulus of a constant radius of $R=12.0 \pm
0.5$ kpc. Figure 12 shows that there is a negative correlation
between the H {\sc i} volume density, $n_{{\rm HI}_0}$, and the scale-height, $z_{1/2}$,
which were determined independently of each other in our analysis.  
The H {\sc i} volume density is approximately expressed by $n_{{\rm HI}0}
= 0.63{\rm exp}(-z_{1/2}/340 \mbox{ pc})$ by a least-squares fitting.

\subsection{Possible Systematic Problems}
In this work, resultant maps were obtained based on some assumptions,
which might potentially cause systematic errors in our results. Possible
systematic problems are described below.

\noindent{1. \it Pure circular rotation}---
Though the distances of H {\sc i} clouds were calculated based on an
assumption that the Galaxy rotates in pure circular rotation,
\citet{bur72} mentioned that the streaming motion predicted by the
density wave theory is not negligible. However, because we are 
interested in the global distribution of the H {\sc i} gas, the effects
due to the streaming motion were not taken into account. Since the
streaming amplitudes were estimated to be between 3 and 8 km s$^{-1}$,
the typical errors in the calculated distances of H {\sc i} clouds were on
the order of 1 kpc. Particularly, in the central region, the
non-circular motion may be large, because the Galaxy is believed to have
a bar whose size is estimated to be about 3 kpc. Hence, the H {\sc i}
distribution in the central 3 kpc may be incredible. 

\noindent{2. \it Dependence on rotation curves}---
Our resultant H {\sc i} map is dependent on the rotation curves and 
there occur two possible systematic errors. 
First is the H {\sc i} disk size. If the rotation velocity becomes 
larger values than the one we used, the H {\sc i} disk will be stretched
outwards.  Second is the H {\sc i} contrast. If a rotation curve has
multiple peaks such as those found in rotation curves of \citet{hon97},
H {\sc i} is concentrated in rings at corresponding radii, because the
H {\sc i} density depends on $|dV_{\rm r}/dr|$ in equation
(\ref{TbtonHI}). We used a smooth rotation curve in order to avoid such
artificial rings.  

\noindent{3. \it Cylindrical rotation}---
We assumed that the gas away from the Galactic plane rotates at the same
speed as the gas in the underlying disk. 
This assumption was often used in former studies \citep{loc84,dip91}. 
However, this assumption may
not be adopted for the high-altitude H {\sc i} gas, because it is known that
the rotation velocity at a large distance from the Galactic plane
becomes slower than that of the gas at underlying disk \citep{swa97}
based on 
                                                                                                                                                                                                                     observations of an external edge-on galaxy, such as NGC 891 in H {\sc i} line. 

\noindent{4. \it Fitting method}---
We determined the H {\sc i} distribution in the inner Galaxy by a fitting
method, as mentioned in subsection \ref{inner}. However, this fitting method cannot
always work precisely. Largely deviated H {\sc i} clouds from the
surrounding gas, often seen in resultant maps, are thus doubtful, while
a global feature is plausible. 

\noindent{5. \it Dependence of adopted functions for
$z$-distribution}---
While the function ${\rm sech}^2(z)$ was adopted in our analysis, the
real gas distribution can not be perfectly represented by it, because the
gas is not isothermal \citep{loc84}. We present H {\sc i} column density
maps derived by using other functions, such as a Gaussian function and exponential
function in figure 12. Though the resultant maps depend
on the 
adopted functions, the global structure is not strongly affected by the 
difference of the chosen functions, but the resultant densities varies
by a factor of $\sim 2$ inside the solar circle.

\section{Summary}
We obtained a three-dimensional H {\sc i} distribution in the Milky Way
Galaxy using the Leiden--Dwingeloo H {\sc i} survey \citep{har97}, Parkes
survey \citep{ker86}, and 
the NRAO survey \citep{bur83} by combining with \citet{cle85} and
\citet{deh98}'s rotation curves. The distance and density of H {\sc i}
clouds were determined by assuming that the 
Galaxy rotates circularly. We solved the near--far problem in the inner
Galaxy using a fitting method based on a model of the vertical
distribution \citep{spi42}. 

Our resultant map shows:   

\noindent{1.} The Galaxy has three prominent spiral arms: the Outer arm,
the Sagittarius--Carina arm, and the Perseus arm. All of the arms are
logarithmic spiral arms.

\noindent{2.} The H {\sc i} disk is strongly warping in the outer disk
and the warping is asymmetric. Also, the inner disk is found to be tilting. 

\noindent{3.} The size of the H {\sc i} disk is about 17 kpc, and the
H {\sc i} mass within this radius is about $2.5 \times 10^9$ $M_\odot$. 

\noindent{4.} The H {\sc i} disk is swelling in the southern hemisphere, which
indicates that the Galaxy is significantly lopsided. 

\noindent{5.} The scale-height increases with the radius globally. The
scale-height is correlated with the H {\sc i} volume density at the centroid
of the H {\sc i} layer. 
 
\bigskip

\noindent{\bf Acknowledgement}
This work was achieved by using the large H {\sc i} survey data of
Leiden-Dwingeloo northern hemisphere survey, Parkes southern hemisphere
survey, and NRAO Galactic center survey. We would like to thank
Dr. T. Sawada, Dr. J. Koda, and Dr. T. Handa for suggestions about writing
the manuscript and for fruitful discussions.









\newpage

\parindent=0pt
\parskip=4mm
\noindent Figure Captions
\vskip 5mm

\noindent{Fig.1.}
Rotation curve used in this paper. The rotational velocity in
 range of $R < 8$ kpc is cited from \citet{cle85}, and that in range of $R > 8$ kpc is cited from \citet{deh98}.

\noindent{Fig.2.}
Radial velocity versus the heliocentric distance in the direction
 of $l=30\arcdeg$ and $b=0\arcdeg$.

\noindent{Fig.3.}
Radial velocity field of the Galactic plane calculated with 
 the rotation curve. The solid lines
 denote positive velocities and the dashed lines denote negative velocities.

\noindent{Fig.4.}
Scale-height versus the Galactocentric distance.

\noindent{Fig.5.}
H {\sc i} column density map. The contours are drawn at levels
 of 8.0e+19, 1.6e+20, 3.2e+20, 6.4e+20, 1.3e+21 
 atom cm$^{-2}$.

\noindent{Fig.6.}
Horizontal cross--section maps. Each map shows the H {\sc i} volume densities
 in a sheet sliced parallel to the Galactic equator. The interval of each
 map is 500 pc; $z=+1000$ pc, $+500$ pc, $0$ pc, $-500$ pc, and $-1000$ pc. 
 The contours are drawn at levels of 0.01, 0.02, 0.04, 0.08, 0.16, 0.32,
 0.64 atom cm$^{-3}$.

\noindent{Fig.7.}
Vertical cross--section maps. Each map shows the H {\sc i}
 volume densities in a sheet through the Galactic center perpendicular
 to the Galactic plane. Individual panel shows sections through
 $\theta=50\arcdeg$ -- $230\arcdeg$, $\theta=80\arcdeg$ -- $260\arcdeg$,
 $\theta=110\arcdeg$ -- $290\arcdeg$, and  $\theta=140\arcdeg$ --
 $320\arcdeg$. The contours are drawn at levels of 0.01, 0.02, 0.04,
 0.08, 0.16, 0.32, 0.64 atom cm$^{-3}$.

\noindent{Fig.8.}
H {\sc i} column density map in polar coordinates. The
 vertical axis denotes the Galactocentric radius on the logarithmic
 scale, and the horizontal axis denotes the azimuthal angle around the
 Galactic center. The contours are drawn at  levels of 1.0e+20,
 2.0e+20, 4.0e+20,8.0e+20, 1.6e+21 atom cm$^{-2}$.

\noindent{Fig.9.}
(a)Schematic tracers of H {\sc i} arms superposed on a resultant H {\sc i}
 column density map. (b) Schematic tracers on a polar--coordinates diagram.

\noindent{Fig. 10.}
Surface density of the H {\sc i} gas versus the Galactocentric distance.

\noindent{Fig. 11.}
Plots of centroid $z_0$ of the H {\sc i} disk versus the
 azimuthal angle $\theta$ around the Galactic center at $R=5$ kpc.

\noindent{Fig. 12.}
Changes in the scale-height at an annulus of constant radius
 $R=12.0 \pm 0.5$ kpc. The horizontal
 axis denotes the H {\sc i} density at the centroid of the H {\sc i} layer
 ($n_{{\rm HI}_0}$) and the vertical axis denotes the scale-height ($z_{1/2}$).

\noindent{Fig. 13.}
(a) H {\sc i} column density map derived by using a Gaussian
 function to determine the H {\sc i} distribution in the inner Galaxy. 
(b) H {\sc i} column density map derived with an exponential function. 
The contours are drawn at levels of 8.0e+19, 1.6e+20, 3.2e+20, 6.4e+20, 1.3e+21 atom cm$^{-2}$.

\end{document}